\documentstyle[aps,pra,graphics]{revtex}

\begin{document}

\draft
\narrowtext
\twocolumn
\wideabs{
\title{Estimating mixed quantum states}

\author{Dietmar G. Fischer and
Matthias Freyberger} 
\address{Abteilung f\"ur Quantenphysik, Universit\"at Ulm, 
D-89069 Ulm, Germany} 


\maketitle
\begin{abstract}
We discuss single adaptive measurements for the estimation of mixed quantum
states of qubits. The results are compared to the optimal estimation schemes
using collective measurements. We also demonstrate that the advantage of
collective measurements increases when the degree of mixing of the quantum
states increases.
\end{abstract}
}

\narrowtext
\section{Introduction}
One of the best distinctions between classical and quantum systems can be
formulated using the language of
state measurements. The state of a single classical system is
an observable: it is in principle always 
possible to measure generalized coordinates
and generalized momenta at a certain time $t$. This information is complete in
the sense that it allows us to calculate the classical state at any time if the
forces are known.\par
Quantum mechanics shows a different picture. In general, the state of a
single quantum system cannot be determined since any measurement will lead to a
reduction of the quantum state. As a consequence the 
complete reconstruction of a
quantum state is only possible if we measure specific observables on an infinite
ensemble of identically prepared systems. Such measurements have been
intensively discussed experimentally as well as theoretically in recent years
\cite{overview}.\par
The interest in such questions has been renewed in the field of quantum
information. In particular, it has been investigated how well one can estimate a
quantum state from a finite ensemble of identical systems. For pure spin states 
this problem has been
solved \cite{massar,derka,tarrach3,bruss}. The intriguing result is that one can
learn more about the finite ensemble by performing a measurement on all quantum
systems simultaneously \cite{peres}. In fact one can optimize the readout of
quantum information via simultaneous strategies. 
Such a joint measurement may, however, pose practical
problems since we may not have all systems available at a time or cannot realize
the required complicated measurement operators experimentally. Hence we have
recently analyzed how close one can come to such optimal joint estimation
schemes by using single adaptive quantum measurements \cite{fischer,mack}.\par
Also for mixed spin states the optimal generalized measurement has been
constructed for a finite ensemble \cite{tarrach}. In addition it has been shown
\cite{vidal} how the gain of information of generalized measurements
increases when the estimated state approaches a pure quantum state. \par
In the present paper we shall analyze the estimation of mixed states of qubits
using
single adaptive measurements. As a criteria for adaption the Kullback
information gain turns out to be suited. We will compare our results to the
optimal measurements discussed in \cite{tarrach}.\par
The paper is organized as follows. In Sec. II we present the quantum system we
are dealing with and introduce the corresponding notation. Sec. III contains the
description of our adaptive measurement scheme. The optimization strategies
used in these adaptive schemes are then described in detail in Sec. IV. In Sec.
V and VI we show the estimation results for the cases of estimation without and
with a priori information regarding the radial distribution of quantum
states inside the Bloch sphere. We conclude with Sec. VII. 
\section{Quantum system and measurement operator}

Suppose we are given
$N$ two-level systems (qubits) 
identically prepared in the mixed state $\hat{\rho}$. The
task is to adaptively estimate $\hat{\rho}$ by using experimentally realizable
single measurements on the two-level systems.\par
Let us first define the notation and the key elements of our analysis.
Any mixed quantum state $\hat{\rho}$ 
of a two-level system (qubit) can be written down in the
Bloch-sphere representation
\begin{equation}
\hat{\rho}(r,\theta,\phi)=\frac{1}{2}\left(\hat{1}+\vec{r}(r,\theta,\phi)
\vec{\hat{\sigma}}\right)
\end{equation}
with the Pauli-spin-vector $\vec{\hat{\sigma}}=(\hat{\sigma}_x,\hat{\sigma}_y,
\hat{\sigma}_z)^T$ and the Bloch vector
\begin{equation}
\vec{r}(r,\theta,\phi)=r\left(
\begin{array}{c}
\sin\theta\cos\phi\\
\sin\theta\sin\phi\\
\cos\theta
\end{array}
\right)
\end{equation}
expressed in spherical coordinates. The parameters $r\in[0,1]$,
$\theta\in[0,\pi]$, $\phi\in[0,2\pi)$ uniquely determine the quantum state
inside
the Bloch sphere. Using the matrix representation of $\vec{\hat{\sigma}}$
 the density
matrix reads
\begin{equation}
\hat{\rho}(r,\theta,\phi)=\frac{1}{2}
\left(
\begin{array}{cc}
1+r\cos\theta & r\sin\theta e^{-i\phi}\\
r\sin\theta e^{i\phi} & 1-r\cos\theta
\end{array}
\right).
\end{equation}
\par
This parametrization also allows us to represent the estimated state 
$\hat{\rho}^{(est)}$ obtained after a certain measurement sequence. By
introducing the probability density $w(r,\theta,\phi)$ we can write  
\begin{equation}
\hat{\rho}^{(est)}=\int dV\; w(r,\theta,\phi)
\hat{\rho}(r,\theta,\phi)
\end{equation}  
with normalization
\begin{equation}
\int dV\; 
w(r,\theta,\phi)=1.
\end{equation}
The integration
\begin{equation}
\int dV=\int^1_0 dr r^2\int^{\pi}_0 d\theta \sin{\theta}\int^{2\pi}_0 d\phi\;
\end{equation}
ranges over the whole sphere.\par
To guarantee the experimental realizability of our proposed measurement strategy
we restrict ourselves to a simple class of measurements.
 This class consists of von Neumann measurements,
e.g., polarization or spin
measurements along a certain axis $(\theta_n,\phi_n)$. 
The tunable parameters $\theta_n$ and $\phi_n$ 
define the direction of the
projection on the Bloch sphere surface for the $n$th measurement.
The corresponding projection 
operator $|\theta_n,\phi_n\rangle\langle\theta_n,\phi_n|$
with the state
\begin{equation}
|\theta_n,\phi_n\rangle=
\cos\frac{\theta_n}{2}|0\rangle+\sin\frac{\theta_n}{2}e^{i\phi_n}
|1\rangle
\end{equation}
therefore defines two measurement results. Either the system described by
$\hat{\rho}(r,\theta,\phi)$ is polarized in the direction $(\theta_n,\phi_n)$
or in the opposite direction given by $(\pi-\theta_n,\pi+\phi_n)$. We encode the
first result by the number 1 and the second by the number 0.\par

The two possible outcomes of the measurement occur with the probabilities
\begin{eqnarray}
P_1(r,\theta,\phi|\theta_n,\phi_n)&=&\langle\theta_n,\phi_n|
\hat{\rho}(r,\theta,\phi)|\theta_n,\phi_n\rangle
\nonumber \\ &=&
\frac{1}{2}\left[1+r\cos\theta\cos\theta_n  \right.  \nonumber \\
 &+& \left. r\sin\theta\sin\theta_n 
\cos(\phi-\phi_n)\right]\nonumber, \\
P_0(r,\theta,\phi|\theta_n,\phi_n)&=&1-P_1(r,\theta,\phi|\theta_n,\phi_n)
\nonumber \\ &=&
\frac{1}{2}\left[1-r\cos\theta\cos\theta_n \right.\nonumber \\
&-& \left. r\sin\theta\sin\theta_n 
\cos(\phi-\phi_n)\right]
\end{eqnarray}
which clearly depend on the chosen measurement direction and on the measured quantum state.

\section{Adaptive measurements}

We now propose an adaptive measurement strategy  
to improve the estimation of a quantum
state 
$\hat{\rho}(R,\Theta,\Phi)$
from a finite ensemble of $N$ identically prepared quantum systems. Note that
the Bloch vector coordinates $R,\Theta$ and $\Phi$ are the same for all $N$ systems.
Despite the fact that we restrict
ourselves to simple projection measurements on single quantum systems 
we will show that it is possible to improve
the estimation quality by using an adaptive 
measurement strategy. This strategy is 
based on an algorithm, cf.
Fig. 1, which consists of five steps:
\begin{enumerate}
\item We take the first, $n=1$, of the $N$ quantum systems and perform a
measurement with randomly chosen direction $(\theta_1,\phi_1)$. 
\item The $n$th measurement along the direction $(\theta_n,\phi_n)$ 
yields one of the two possible outcomes that we denote by
$i=0$ and $i=1$. 
By using this information we modify the distribution
$w_{n-1}(r,\theta,\phi)$ of the estimated density operator
\begin{equation}
\hat{\rho}_{n-1}^{(est)}=\int dV\; 
w_{n-1}(r,\theta,\phi)\hat{\rho}(r,\theta,\phi)
\end{equation}
after $n-1$ steps
according to Bayes' rule \cite{bayes}
\begin{equation}
w_{n}(r,\theta,\phi)=Z^{-1}
P_i(r,\theta,\phi|\theta_n,\phi_n)w_{n-1}(r,\theta,\phi).
\end{equation}
Hence with the help of the probabilities $P_i$, Eq. (8), we update our current
knowledge about the finite ensemble as formulated by the distribution $w_n$.
The normalization constant reads 
\begin{eqnarray}
Z=\int dV\; P_i(r,\theta,\phi|\theta_n,\phi_n)w_{n-1}(r,\theta,\phi). &&
\end{eqnarray}
Before we have acquired any information about
the system, we assume our knowledge to be homogeneously distributed over the Bloch
sphere; that is, we start from the initial distribution 
\begin{equation}
w_0(r,\theta,\phi)=\frac{3}{4\pi},
\end{equation}
thereby assuming that each volume element $dV=r^2\sin\theta dr d\theta d\phi$
is equally probable.
\item  The updated probability distribution $w_{n}(r,\theta,\phi)$ describes our
present knowledge about the quantum state. With its help we determine the
next measuring operator, i.e., the direction $(\theta_{n+1},\phi_{n+1})$ of the measurement step 
$n+1$. The new measuring operator should be designed in such a way that
it allows us to gain the maximum amount of additional information about the unknown quantum
state $\hat{\rho}$. For this purpose we have to apply a criterion by which we 
quantify the
notion of maximum information gain. The different criterions that we use in this
context will be described in detail in the next section. This step reflects the 
adaptive aspect of our
algorithm, because the choice of a measuring operator is based on 
$\hat{\rho}_n$ and thereby on the history of all previous measurement outcomes. 
\item Once having found the next measuring operator 
$|\theta_{n+1},\phi_{n+1}\rangle\langle\theta_{n+1},\phi_{n+1}|$ 
we now take one of the
remaining quantum
systems and measure it. If we still have quantum systems
left, we continue with step 2. 
\item After we have used up all $N$ mixed qubits we arrive at the final probability
distribution $w_N(r,\theta,\phi)$ which allows us to construct 
the corresponding estimated state 
$\hat{\rho}_N^{(est)}$.
\end{enumerate}
As the measure of our
state estimation quality we will use the 
fidelity\cite{uhlmann,josza} 
\begin{equation}
F_N\left(\hat{\rho},\hat{\rho}_N^{(est)} \right)=
Tr^2\sqrt{\sqrt{\hat{\rho}_N^{(est)}}\hat{\rho}(R,\Theta,\Phi)\sqrt{\hat{\rho}_N^{(est)}}},
\end{equation}
for mixed quantum states which reduces to\cite{josza,huebner}
\begin{eqnarray}
F_N=&\frac{1}{2}&\left[1+\vec{r}_N^{(est)}\vec{r}(R,\Theta,\Phi)
\right. \nonumber \\ &+& \left. \sqrt{1-|\vec{r}_N^{(est)}|^2} 
\sqrt{1-|\vec{r}(R,\Theta,\Phi)|^2}\right]
\end{eqnarray}
for two-level systems with Bloch vectors $\vec{r}_N^{(est)}$ and $\vec{r}(R,\Theta,\Phi)$.
Note that this fidelity of course depends on the number $N$ of quantum systems at our
disposal.

\section{Measurement Strategies}
In this section we will describe the strategies that we have 
applied to
find the direction $(\theta_n,\phi_n)$ for the $n$th measurement by learning from the results
of earlier measurements.

\subsection{Random selection from all axes}
A straightforward way to select a new measurement direction 
$(\theta_n,\phi_n)$ is to
choose the parameters $\theta_n$ and $\phi_n$ randomly on the Bloch sphere, 
independent of any
knowledge already acquired about the state. That is, each infinitesimal surface element
$\sin\theta_n d\theta_n d\phi_n$ occurs with the same 
probability
$1/4\pi$. Clearly this strategy is not adaptive because 
$(\theta_n,\phi_n)$ does not depend on any previous 
measurement outcome. 
Nevertheless, the estimated density operator
$\hat{\rho}_{n-1}^{(est)}$ can still be updated after each measurement as described in 
step 2
of our algorithm. \par
The random selection implements a measurement protocol lacking any
constructive strategy. Thus the results of this method will serve
as a reference to which we can compare the outcomes of the 
optimization strategies
described below. \par 

\subsection{Minimal measurements along three axes}
In principle it is possible to determine any mixed state by measurements along three
axes if an infinite number of quantum systems prepared in this state is
available. Without loss of generality one can choose the Pauli operators 
$\hat{\sigma}_x$, $\hat{\sigma}_y$ and $\hat{\sigma}_z$ to
represent measurements along three orthogonal 
directions on the Bloch sphere. 
This set of measurement operators represents a minimal measurement or -- in
other words -- corresponds to a minimal quorum. \par
Thus it is interesting to compare the efficiency of such a minimal measurement to our
adaptive methods in the case of a finite number $N$ of available quantum systems.
Then the measurement scheme consists of projecting an average number of $N/3$ systems using each
of the directions $(\pi/2,0)$, $(\pi/2,\pi/2)$ and $(0,0)$.

\subsection{Maximization of Kullback information gain}
If we look at the measurement procedure from an information theoretic point of
view then our aim will be to maximize the  
information that we can get
in the measurement step from $n-1$ to $n$.
A measure for the average information gain in the $n$th measurement 
is the so called Kullback information\cite{vidal,kullback} that can
be defined as
\begin{eqnarray}
\bar{K}(\theta_n,\phi_n)=
&\sum_{i=0}^1& p_i^{(est)}(\theta_n,\phi_n)\nonumber \\
&\times &\int dV\; w_n^{(i)}(r,\theta,\phi)\mbox{ log}_2\;\frac{w_n^{(i)}(r,\theta,\phi)}
{w_{n-1}(r,\theta,\phi)}
\end{eqnarray}
with 
\begin{equation}
p_1^{(est)}(\theta_n,\phi_n)=\langle\theta_n,\phi_n|\hat{\rho}_{n-1}^{(est)}
|\theta_n,\phi_n\rangle
\end{equation}
and
\begin{equation}
p_0^{(est)}=1-p_1^{(est)}
\end{equation} 
being the estimated probabilities for the outcomes $i=0,1$ based on our current knowledge,
i.e., based on the density operator $\hat{\rho}_{n-1}^{(est)}$. Consequently, also the
probability density
\begin{equation}
w_n^{(i)}=Z^{-1}p_i^{(est)}(\theta_n,\phi_n)w_{n-1}
\end{equation}
explicitly depends on the outcome $i$ and on the 
direction $(\theta_n,\phi_n)$. Hence our expression,
Eq. (15), for the
Kullback information is a function of $(\theta_n,\phi_n)$ which can be maximized. In order to
see that $\bar{K}$ describes an estimated average information gain we rewrite it in the form
\begin{eqnarray}
\bar{K}(\theta_n,\phi_n)&=&\sum_{i=0}^1 p_i^{(est)}(\theta_n,\phi_n)
\nonumber \\ &\times &
\int dV\; w_n^{(i)}(r,\theta,\phi)\mbox{ log}_2\; w_n^{(i)}(r,\theta,\phi)\nonumber \\
&-& \int dV
\; w_{n-1}(r,\theta,\phi)\mbox{ log}_2\; w_{n-1}(r,\theta,\phi)\nonumber \\
&=& S_{n-1}-\sum_{i=0}^1 p_i^{(est)}(\theta_n,\phi_n) S_n^{(i)}(\theta_n,\phi_n).
\end{eqnarray}
The entropy $S_{n-1}$ describes our knowledge before the measurement, whereas the entropy
$S_n^{(i)}$ stands for the estimated entropy provided we find the result $i$.\par
Hence we can maximize the difference of entropies before and after the measurement 
by adjusting
the parameters $(\theta_n,\phi_n)$. We therefore select the measuring direction 
$(\theta_{n},\phi_{n})$ that yields the maximum average information gain.
\par
The estimated state $\hat{\rho}_N^{(est)}$ is 
finally again 
determined from $w_N$. This strategy can be applied in the case that all
measurement directions are possible as well as in the case that only the three
measurement directions along 
$\hat{\sigma}_x$, $\hat{\sigma}_y$ and $\hat{\sigma}_z$ are allowed. In the
latter case the maximization described before is done only for the directions 
$(\pi/2,0)$, $(\pi/2,\pi/2)$ and $(0,0)$.
We will
discuss the resulting estimation precisions in the next section.
 
\section{Estimation without \it a priori \bf information}
In this section we numerically evaluate the average fidelities for the
state estimation schemes described above. Only such an average fidelity is a reasonable
measure of quality of a specific estimation procedure, since we assume to
have no prior information about $\hat{\rho}$.\par
These average fidelities
will depend on the  
number $N$ of identically prepared quantum systems that we have at our disposal.
Thus one state estimation experiment
consists of a sequence of $N$ measurements performed on $N$ identical systems in
state $\hat{\rho}(R,\Theta,\Phi)$ and a subsequent estimation of a mixed
state
$\hat{\rho}_N^{(est)}$. The fidelity $F_N$, Eq. (14), 
of the state
estimation is then calculated by comparing both states.\par
However, in order to get the average fidelity 
\begin{equation}
\big{\langle}F_N\big{\rangle}=\Big{\langle}F\left(\hat{\rho}(R,\Theta,\Phi),
\hat{\rho}_N^{(est)}\right)\Big{\rangle}_{\hat{\rho}}
\end{equation}
we have to perform such a single run of the (numerical) 
experiment over and over again for different
initial states $\hat{\rho}(R,\Theta,\Phi)$, i.e., for different coordinates 
$(R,\Theta,\Phi)$. 
Hence the initial states $\hat{\rho}(R,\Theta,\Phi)$, Eq. (3), 
are chosen randomly from an isotropic and homogenous probability
distribution $3/(4\pi)$ for each volume element $R^2\sin\Theta dR d\Theta
d\Phi$ of the Bloch sphere. 
Using this
homogenous probability distribution ensures that the performance of an estimation strategy 
is not biased by any specific choice of initial states. For more details on the averaging
procedure see Appendix A.\par
The average
fidelity for each $N$ was obtained by 
averaging over $10^4$ experiments, i.e., $10^4$ initial states equally
distributed inside the Bloch sphere. 
For the sake of a clear graphical presentation
not the average fidelities
themselves but the average errors
\begin{equation}
f_N=1-\big{\langle}F_N\big{\rangle}
\end{equation}
are calculated for different $N$. \par 
In Fig.2 the average errors 
$f_N$ are compared to the average error 
$f_N^{rand}$
of the
random selection scheme by plotting the ratio
\begin{equation}
\gamma_N\equiv\frac{f_N}{f_N^{rand}} 
\end{equation}
versus $N$. This quantity shows the relative performance of the different
schemes compared to the random selection scheme\cite{precision}.\par
As we can see the 
scheme based on measurements along three axes improves the estimation quality
by approximately 
4\% even if no adaptive strategy is applied. We can further decrease
the estimation error if we use the 
Kullback information gain strategy.
For this strategy the average errors are always smaller than for the non-adaptive
schemes in both cases. This
shows that one can indeed decrease the error of a state estimation by using
adaptive algorithms. The errors decrease by approximately 3\% for $N>10$ in
the case of the all-axes scheme. And even in the case of only three
possible measurement directions an optimization of the Kullback information gain
yields an improvement of the estimation quality. Moreover, the resulting measurement
strategy is the best strategy based on separate measurements that we found so
far. \par
Now we analyze how our estimation schemes compare to the optimal
ones \cite{tarrach} that are based on collective measurements on all $N$ quantum
systems. For pure qubits it was shown \cite{fischer} that the fidelities of
single qubit measurement schemes can be always bigger than 98\% of the optimal
ones \cite{massar,derka,tarrach3} 
that use collective measurements. Can we also reach
such values in the case of mixed qubits? It turns out that this is not the
case. In Fig. 3 we show the ratios $\big<F_N\big>/\big<F_N^{(opt)}\big>$ 
with $\big<F_N^{(opt)}\big>$ being
the fidelity of the optimal measurement scheme \cite{tarrach} 
for different $N$. We find that for the random selection scheme we get
fidelities which are in the worst case below 96\% of the optimal values. Even
for the best estimation scheme (measurement along three axes with Kullback
information maximization) we are still about 3.5\% off the optimal values. Also
the convergence towards 1 is much slower for mixed qubits than in the case of
pure ones \cite{fischer}. For mixed qubits we are still about 2\% off the
optimal fidelities at $N=50$ whereas for pure qubits the ratio of fidelities has
already been very close to 1 at this point.\par
These results indicate that the advantage of collective measurements compared to
single quantum system measurements increases with growing degree of mixing of
the qubits. Our results therefore confirm the asymptotic findings by Gill and
Massar\cite{massar2} who investigated the case of state estimation of large
qubit ensembles. They showed that for large $N$ one can asymtotically achieve
the precision of collective measurement schemes by measurements on single
qubits if they are known to be in a pure state and that this is no longer true
in the case of mixed qubits. 

\section{Estimation with radial \it a priori \bf information}

Inspired by these findings we finally discuss how the precision of our estimation
schemes depends on the degree of mixing of the initial states.
In our case the adequate measure of precision is
the estimation error $f_N$, Eq. (21). For a quantitative check of the relation
between estimation quality and degree of mixing of the initial states we have
chosen the initial states no longer according to a homogenous probability
distribution inside the Bloch sphere but from a probability distribution 
\begin{equation}
w_0(r,\theta,\phi;\alpha)dV=\frac{\alpha+1}{4\pi}r^{\alpha}\sin\theta dr d\theta
d\phi
\end{equation}
which clearly depends on the radius $r$. Please note that we get the 
homogenous distribution for $\alpha=2$ again. A larger parameter $\alpha\ge 0$ 
means
a growing average radius
\begin{equation}
\bar{r}=\frac{\alpha+1}{\alpha+2}
\end{equation}
which indicates a decreasing degree of mixing of the initial states. Thus a
variation of the parameter $\alpha$ allows us to study the influence of the
degree of mixing onto the estimation precision of our schemes in which we have
of course also adapted the initial probability $w_0$. \par
However, our previously desribed estimation scheme does not converge towards the
estimation scheme for pure states \cite{fischer} for $\alpha\to\infty$. The
cause of this behaviour is the final readout of the estimated radius
$r^{(est)}$. Please remember that this radius is found via the integration $\int
dV w_N(r,\theta,\phi)\hat{\rho}(r,\theta,\phi)$ which yields the estimated state
$\hat{\rho}^{(est)}(r^{(est)},\theta^{(est)},\phi^{(est)})$. Thus even for solely
pure initial states the estimated state will never be pure itself. That is, the
readout of the estimated state does not properly account for our a priori
information. For larger $\alpha$
the estimated radius will always be too small compared to the average radius
$\bar{r}$, Eq. (24), of our initial distribution $w_0$, Eq. (23). \par
To overcome this non-convergence we now introduce an alternative readout scheme
for the radius $r^{(est)}$. The estimated parameters $\theta^{(est)}$ and 
$\phi^{(est)}$ are still obtained in the same way as before, that is, via Bloch
representation of $\hat{\rho}^{(est)}$. The
estimated radius $r^{(est)}$, however, is now chosen as the average radius 
\begin{equation}
r^{(est)}=\int dV\; r\; w_N(r,\theta,\phi)
\end{equation}
of the final distribution $w_N(r,\theta,\phi)$. 
It is easy to show that --- even before the first measurement --- this estimated
radius tends to 1 for large $\alpha$ and merges into the pure state estimation
scheme for $\alpha\to\infty$\cite{scheme}. \par
Using this readout scheme we compare
the resulting errors $f_N$ to the errors $f_N^{(opt)}$ that could be achieved by a optimal
collective estimation scheme \cite{tarrach}. 
The errors $f_N$ are based on a simple 3-axes estimation as described in Sec.
IV.

In Fig. 4 we plotted the ratio
$f_N^{(opt)}/f_N$ versus $N$. As one would expect our estimations are always worse
than the optimal ones. We also find a very clear dependence of the ratio on the
radial distribution described by the parameter $\alpha$: The smaller 
$\alpha$ the smaller is the ratio and vice versa. This means that for highly
mixed states the optimal collective measurements offer a bigger advantage
compared to separate measurements than for states with a small degree of mixing.
Or, in other words, the bigger $\bar{r}$, Eq. (24), the closer can we come to the
optimal limits using separate measurements. It is interesting to note that this
statement is true for all $N$ and not only in the limit $N\to\infty$ as shown in
\cite{massar2}.

\section{Conclusion}
We have presented estimation methods
based on separate adaptive measurements on single qubits.
We have demonstrated the measurement schemes for estimating mixed 
quantum states of qubits. 
An algorithm is used to update the
knowledge about the true quantum state after each measurement and to choose the
best measuring operator for the next measurement. With this scheme we have been
able
to reduce the estimation errors compared to non-adaptive strategies. The
best results can be obtained by using schemes related to Kullback information
measures. Maximizing this information gain leads to considerable improvements in
the estimation quality. \par
We have also shown that the advantage of collective
measurements decreases with decreasing degree of mixing of the initial qubits
for all $N$, thereby confirming asymptotic results for $N\to\infty$ found by
Gill and Massar\cite{massar2}.\par
We have restricted ourselves to simple separate measurements which can
be easily realized with nowadays technology.
 An additional
advantage of our scheme is that there is no need to have all $N$ quantum systems
available at the same time. In contrast to optimal measurement schemes, for
which one needs to perform complicated collective measurements on all the
systems,  
our schemes can also be used if the $N$ quantum systems can only be prepared one after
the other. These features ensure the
applicability of our scheme to experiments and practical state estimation
problems in quantum information theory.

\acknowledgments
We acknowledge support by the DFG programme
``Quanten-Informationsverarbeitung'',
by the European Science Foundation QIT programme and by the IST programme
``QUBITS'' of the European Commission. 

\begin{appendix}
\section{Averaging over the Bloch sphere}
In order to quantify the performance of our adaptive methods we have
introduced the average fidelity $\big\langle F_N\big\rangle $, Eq. (20). 
In this Appendix we shortly
describe the averaging procedure. In principle the calculation of $\big\langle F_N\big\rangle $
consists of two steps. First we have to determine the average fidelity
$\bar{F}_N(\hat{\rho})$ for $N$ identical quantum systems prepared in state
$\hat{\rho}=\hat{\rho}(R,\Theta,\Phi)$ by summing over all possible measurement paths $J$.
Given a specific adaptive method each path is uniquely determined by the
initial measurement direction $(\theta_1,\phi_1)$ and by the sequence of
measurement results forming a string of 0's and 1's. That is, we have $J\equiv
J(\theta_1,\phi_1;\{0,1\}^N)$. With the fidelity $F_N(\hat{\rho},J)$ for each path we
arrive at 
\begin{equation}
\bar{F}_N(\hat{\rho})\equiv
\bar{F}_N(R,\Theta,\Phi)=\big\langle F_N(\hat{\rho}(R,\Theta,\Phi),J)\big\rangle_J.
\end{equation}  
In principle a
simulation of this expression would be straightforward. We choose an initial
measuring direction and perform a Monte-Carlo simulation with sufficiently many
measurement sequences of length $N$. This should be repeated for a dense set of
initial measurement directions on the Bloch sphere.  \par
The second step of our averaging procedure consists of averaging
$\bar{F}_N(R,\Theta,\Phi)$ over all density operators $\hat{\rho}=\hat{\rho}(R,\Theta,\Phi)$
isotropically distributed over the Bloch sphere. In this way we find the fidelity
\begin{equation}
\big\langle F_N\big\rangle =\big\langle \bar{F}_N(R,\Theta,\Phi)\big\rangle_{(R,\Theta,\Phi)} 
\end{equation}
which is not biased by any specific choice of density operator and therefore
measures the performance of any estimation method.\par
However, this averaging procedure can be simplified, if we take the following
into account. The fidelity $F_N(\hat{\rho},J)$ is rotationally invariant. For a
density operator rotated on the Bloch sphere with unitary transformation $U$ we
have 
\begin{equation}
F_N(U\hat{\rho} U^{\dagger},J)=F_N(\hat{\rho},U^{\dagger}J U)
\end{equation}
where $U^{\dagger}J U$ symbolizes the corresponding rotated path. Hence
instead of averaging over all paths $J$ in Eq. (A1) we can average over all possible
density operators $\hat{\rho}=\hat{\rho}(R,\Theta,\Phi)$ for a fixed radius $R$, that is 
\begin{equation}
\big\langle F_N(\hat{\rho}(R,\Theta,\Phi),J)\big\rangle _J=\big\langle F_N(\hat{\rho}(R,\Theta,\Phi),J)\big\rangle_{(\Theta,\Phi)}=
\bar{F}_N(R)
\end{equation}
with $J$ chosen randomly for each setting $(\Theta,\Phi)$. Note that this also
reduces $\bar{F}_N$ to a pure function of the radius $R$. Consequently, the
final fidelity reads
\begin{equation}
\big\langle F_N\big\rangle =\big\langle \bar{F}_N(R)\big\rangle _R=\big\langle F_N(\hat{\rho}(R,\Theta,\Phi),J)\big\rangle_{(R,\Theta,\Phi)}.
\end{equation}
Therefore, we numerically simulate $\big\langle F_N\big\rangle $ by choosing sufficiently many points
$(R,\Theta,\Phi)$ isotropically distributed over the Bloch sphere together with
a randomly chosen path $J$.

\end{appendix}



\begin{figure}
\resizebox{8cm}{!}{\includegraphics[95,200][550,600]{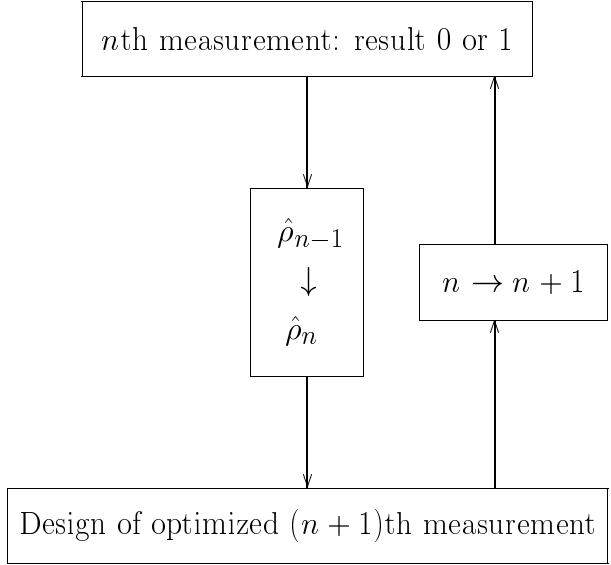}}
\caption{Schematic picture showing the sequence of steps in the adaptive
algorithm. The algorithm starts with performing the $n$th experiment. The
measurement result is then used to update the estimated density operator
$\hat{\rho}_{n-1}$. With the help of the updated density operator 
$\hat{\rho}_{n}$ the measuring operator for the $(n+1)$th measurement is 
selected and the algorithm restarts.}
 
\end{figure}

\begin{figure}
\rotatebox{-90}{\resizebox{6cm}{!}
{\includegraphics[95,140][550,600]{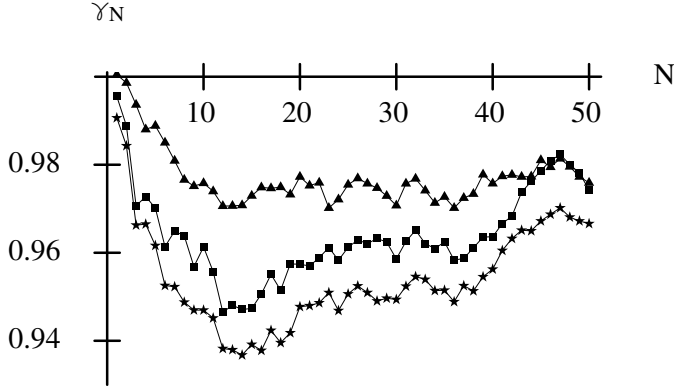}}}
\caption{Relative error $\gamma_N$, Eq. (22), plotted versus number of quantum systems
$N$. The boxes show the results for the 3-axes measurement scheme,
whereas the triangles (stars) visualize the relative errors of the
Kullback information gain strategies without (with) restriction to only three
possible measurement axes.}
\end{figure} 
\begin{figure}
\rotatebox{-90}{\resizebox{6cm}{!}
{\includegraphics[95,100][550,600]{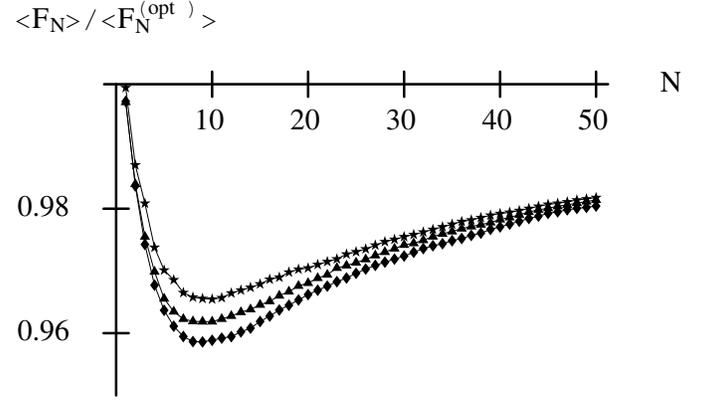}}}
\caption{Ratio of average fidelity $\big<F\big>$ and optimal average 
fidelity $\big<F_{opt}\big>$ plotted versus
$N$. 
Triangles (stars) again
describe the adaptive strategies without (with) restriction to three axes.
In addition the ratio for the random selection scheme is also plotted
(diamonds).}
\end{figure}
\begin{figure}
\rotatebox{-90}{\resizebox{6cm}{!}
{\includegraphics[95,100][550,600]{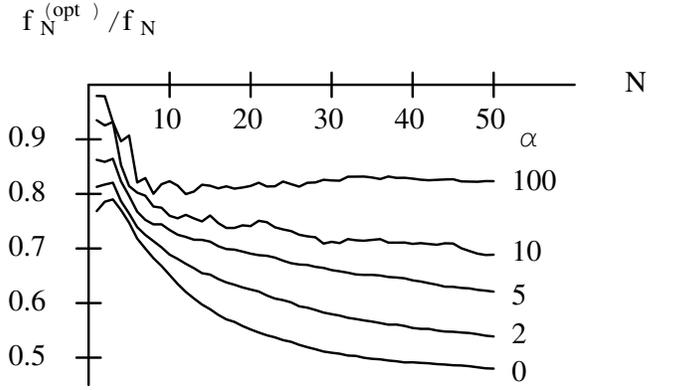}}}
\caption{Ratio of optimal estimation error $f_N^{(opt)}$ and error of 3-axes
estimation scheme $f_N$ plotted versus $N$ for diffferent radial distributions of
initial states parametrized by $\alpha$, Eq. (23). From bottom to top 
the lines correspond
to $\alpha=0$, 2, 5, 10, 100. We clearly see an increasing ratio for growing
$\alpha$.}  
\end{figure}


\begin{thebibliography}{99}
\bibitem{precision} The resulting numerical accuracy of the mean errors is
approximately 1\% of the error values themselves.
\bibitem{scheme} The reason why we do not use this readout
scheme for a homogeneous probability distribution ($\alpha=2$) is that our
original readout scheme yields better results in this case, especially for
small $N$. Only for larger $\alpha$, i.e., for more a priori knowledge, 
the readout of $r^{(est)}$ via Eq. (25) 
becomes superior.
\end{thebibliography}
\end{document}